\documentclass[twocolumn,10pt]{IEEEtran}
\usepackage{url}
\usepackage{amsmath}
\usepackage{amsfonts}
\usepackage{amssymb}
\usepackage{mathrsfs}
\usepackage{algorithm}
\usepackage{algorithmic}
\usepackage{pifont}
\usepackage{float}
\usepackage{upgreek}
\usepackage{color}
\usepackage[final]{graphicx}
\usepackage{caption}
\usepackage{epsfig}
\usepackage{rotating}
\usepackage{physics}
\usepackage{mathtools}
\usepackage{cuted}

\usepackage[numbers,sort&compress]{natbib}
\bibpunct[, ]{[}{]}{,}{n}{,}{,}

\newcommand*{\im}{{\mathrm{i}}}

\begin{document}
\title{Stability tests in time of OAM multiplexing schemes in highly disturbed environments.}
\author{\IEEEauthorblockN{F. Tamburini\IEEEauthorrefmark{1}, B.\,Thid\'e\IEEEauthorrefmark{2},
 P. Weibel\IEEEauthorrefmark{1}, V. Boaga\IEEEauthorrefmark{3}, F. Carraro\IEEEauthorrefmark{3}, and M. Del Pup\IEEEauthorrefmark{3}.
}
\\
\IEEEauthorblockA{\IEEEauthorrefmark{1} ZKM -- Zentrum f\"ur Kunst und Medientechnologie, Lorentzstr. 19, D-76135, Karlsruhe, Germany, EU}
\\
\IEEEauthorblockA{\IEEEauthorrefmark{2} Swedish Institute of Space Physics, {\AA}ngstr\"{o}m Laboratory, P.~O.~Box~537, SE-75121 Uppsala, Sweden, EU}
\\
\IEEEauthorblockA{\IEEEauthorrefmark{3} ARI, Sezione Venezia, Santa Croce~1776/B, IT-30123 Venice, Italy, EU}
\thanks{Corresponding author F. Tamburini. Email: fabrizio.tamburini@gmail.com}}
%


\maketitle

\begin{abstract}
We report the results of tests of data transmission and signal stability in time of two different wide-band multiplexing (MUX) schemes, each in a point-to-point configuration, based on electromagnetic waves carrying Orbital Angular Momentum (OAM) in noisy real-world settings.
Each radio link transmitted two high definition wide--band analog TV channels in the same frequency band with FM-carrier centered at $2.414$ GHz and $27$ MHz bandwidth, encoded with different OAM modes in the same polarization state, uninterruptedly for $5$ months during the world exhibition ``Globale--Digitale'' at ZKM in Karlsruhe and in other $2$ months time slots taken in the following $4$ years, $24$ hours per day.
We show the practical feasibility of the use of stable OAM radio/TV links in the real world for a long time, paving the way for for secure and efficient communication schemes also under electromagnetic jamming conditions.
\end{abstract}


\section{Introduction}
\IEEEPARstart{C}{Urrent} radio science and communication implementations based on the electromagnetic (EM) linear momentum $\mathbf{P}$ physical layer, associated with the Poynting vector and the translational dynamics and force action of the system, are approaching their limits in terms of radio frequency spectrum availability and occupancy even with the use of multiple-input multiple-output (MIMO) techniques \cite{mimo} and the additional use of wave polarization that can at most double the channel capacity within a given oscillation frequency bandwidth.
This calls for the introduction of new radio paradigms to transfer more information, namely, to use not only the energy $\mathbf{E}$ and the linear momentum $\mathbf{P}$ conserved quantities, but also the other existing partial and total conserved quantities of the EM field \cite{fuschichnikitin} still underutilized in radio science and technology.
For this purpose, it has been proposed that the EM angular momentum conserved quantity $\mathbf{J}$, associated with rotational dynamics and torque action of the system, as described in the standard literature
\cite{Heitler1954,Podolsky&Kunz1969,Eyges1972,Berestetskii&al1980,Ribaric&Sustersic1990,Mandel&Wolf1995,Schwinger&al1998,Jackson1998,Allen&al2003,Rohrlich2007,Andrews2008,Torres&Torner2011,Yao&Padgett2011,Thide2011}, could be fully exploited in radio communications to improve the existing multiplexing systems 
\cite{Thide&al:PRL:2007,Franke-Arnold&al:LPR:2008,Tamburini2011,Tamburini&al:APL:2011,Tamburini&al:NJP:2012,Tamburini&al:NJP:2012a,thidepreprint2014}.

The total EM angular momentum conserved quantity $\mathbf{J}$ can be split in two sub-components: the spin angular momentum (SAM)  $\mathbf{S}$, associated with the polarization of the EM wave and the spin of a photon, and the orbital angular momentum (OAM),  $\mathbf{L}$ \cite{Torres&Torner2011,Thide2011}.
SAM and OAM cannot in general be considered as separate quantities that evolve independently during the evolution of the system. This can happen only in the paraxial approximation when propagating in free space or in a homogeneous isotropic medium.

As follows from Maxwell's equations, OAM is a pseudovector, associated with a variation of the phase of the 
EM field as a function of the azimuthal angle, $\varphi$, taken in the plane perpendicular to that of the wave's propagation and centered on the axis of symmetry of the beam.
OAM is related to the rotations in space around a given axis i.e. with recurrence modulo $2 \pi$. Because of this, OAM takes discrete values. It is expressed as a discretized quantity even in a classical framework where a macroscopic EM field is present. 
Any EM field, carrying an arbitrary amount of angular momentum, can be expanded in a superposition of a denumerably infinite set of discrete OAM modes. An example are Laguerre-Gaussian (LG) modes that present a cyclic phase factor $\exp{(\im m\varphi)}$ modulo the integer $m=0, \, \pm1, \pm2, \ldots$, also known as topological charge \cite{Allen&al2003,Andrews:Book:2008,Torres&Torner2011}.
In these beams each photon carries a well-defined quantity $m \hbar$ of OAM. A plane wave does not carry orbital angular momentum and has $m=0$ OAM.
OAM modes with different values of $m$ are mutually orthogonal and are therefore independent even when they have the same $m$ value with opposite signs. It is interesting to point out that OAM is a pseudovector and, because of this, any reflection will change the sign of the OAM and the reflected beam will result orthogonal to the original one.

OAM, together with SAM, represents a new physical layer for radio science and technology. In the paraxial approximation, when propagating in free space, SAM and OAM can be used to set up efficient communication schemes that can be in certain cases equivalent or even complementary to the already existing MIMO systems \cite{oldoni}. 
Different signals, encoded in different OAM states independently generated in the same frequency band, do not interfere with each other \cite{Molina-Terriza&al:PRL:2002} and can therefore be used to encode in a more efficient way a larger amount of information down to the single photon level \cite{Mair&al:N:2001}. 
This has been confirmed by numerical simulations, experiments \cite{Thide&al:PRL:2007,Tamburini2011} in controlled anechoic chamber \cite{Tamburini&al:APL:2011} and outdoor in a real-world setting in the radio \cite{Tamburini&al:NJP:2012,Tamburini&al:NJP:2012a,Tamburini&al:preprint:2013a} and optical bands \cite{Gibson&al:OE:2004,Celechovsky&Bouchal:NJP:2007,Wang&al:NPHO:2012} that gave the start to the vast literature present nowadays.

One of the most important properties of OAM modes is that they are physical states of the EM field that preserve their topological properties of phase and intensity during their propagation in free space and represent a viable solution for a stable multiplexing (MUX) realized with standard linear momentum--based techniques \cite{bilotti18} and the physical properties of the OAM field \cite{wileyIEEE,amradio1,amradio2}.
The robustness of the OAM beams during their propagation is confirmed by the results of a recent work where the topological stability of radio vortices at $233$GHz permitted to measure the rotation of the supermassive black hole in the galaxy M87, after having traveled a distance of $56$ millions of light years in the outer space \cite{Tamburini2011,tamburinimnras2019,FFT}.

In this paper, we report the results of the tests for the stability in time of two OAM multiplexing systems for wide-band radio/TV transmission indoor and the robustness with respect to the presence of multiple reflections and disturbing signals. The two experiments ran uninterruptedly $24$ hours per day. The longer period was $5-$months long in 2015 during the Globale--Digitale. Additional runs were made every year for 2 months including additional tests with digital channels.
This represents a common real-world situation that can be found in MUX TV links indoor.
\section{Experimental results}

\textbf{Channel stability in time:} The two OAM links -- experiment 1 and experiment 2 -- were part of scientific and artistic installations present in the 2015 Globale-Digitale world exhibition in Karlsruhe,
``The New Art Event in the Digital Age'' took place over 300 days in Karlsruhe, where ZKM presented spectacular installations, innovative works of art at the interface with natural science, science with scientific experiments as new form of science in terms of a new expression of art, urban area performances accompanied by concerts, lectures and conferences \cite{globale}.
As part of Globale, the ``Infosphere'' exhibition presented an overview of art in the era of the digital revolution, with science and its social consequences, giving a new perspective to the new data and science worlds \cite{Infosphere}.

Both the links ran uninterruptedly for $5$ months $24$ hours per day with more than $20.000$ people assisting at the conferences and exhibitions interacted with the experiments and $2$ months per year in the following years until $2019$ in similar conditions. This replicates exactly the conditions in which OAM links would experience when applied to real world settings where random disturbing radio signals were generated by the local wifi, bluetooth systems and mobile phones of the visitors. Additional static interferences were caused by the reflections from the walls and the metal structures there present.
The installations were mounted in large accessible closed spaces under safe conditions for people.

In both the two experiments, two different and independent links were simultaneously transmitting two superimposed high quality (HD standard) audio and video analog channels on the same frequency band. One channel was twisted, carrying $m=1$ orbital angular momentum, whilst the other one was untwisted ($m=0$). Each dual link transmitted two FM wide-band TV channels in the same frequency band centered at $2.414$ GHz and $27$ MHz bandwidth, one twisted with $m=1$ OAM and the other untwisted ($m=0$). 

We used horizontal polarization for both the links: in experiment 1 to reduce the effects of reflections from the vertical metal structures present in the Foyer of ZKM, a large closed environment. In the second link (experiment 2), set in the ``Infosphere'' zone of the exhibition, the effect of multiple reflections and high electromagnetic pollution due to the presence of other artistic/scientific installations reduced more the channel separation than in experiment 1.
The fading caused by reflections is, in fact, one of the most common complications in wireless communications that can decrease the quality of the transmissions, including those based on multi-path protocols like MIMO (Multiple-Input-Multiple-Output) \cite{mimo}. 
The main requirement for a good quality HD video transmission was to obtain a channel separation with carrier-to-noise ratios of at least $20$ dB in order to have a stable and good quality in both video streamings (twisted and untwisted). 
We periodically measured the channel separation, bit transfer rate and errors with a digital transmission on the twisted channel.
No macroscopic effects due to interferences have been reported during the long runs and no evident signal degradation was observed below the lower limit of $20$ dB required. Down to that value, the presence of continuous or sporadic interferences would have affected the audio-video streaming with clearly evident effects such as the rising of patterning on the video, with lines and/or ``snow'' until the complete blocking of the signal with a frozen image displayed on the screen by the receiver. The interferences from the Wi-Fi, instead, can cause horizontal noise bars on the screen, with additional noise in the audio channel. 

\textbf{Experiment 1:} in the foyer of ZKM in Karlsruhe, always crowded with people, we set up the links with maximum emitting power of $0.1$W for each channel.
The distance between the source and the horizontal bar where the two interferometers were mounted was $17.8$ metres, equivalent to $\sim 145~\lambda$.
The two channels transmitted two high quality movies with artistic and scientific contents that were displayed in two separate screens (more information can be found in Ref. \cite{einsteinsdream}).

The twisted beam was generated by the helicoidally-deformed parabolic antenna used for the $442$-metres link in the Venice experiment \cite{Tamburini&al:NJP:2012}, with $80$~cm ($6.4 \lambda$) diameter, whilst the untwisted beam was generated by a dipole with a $\lambda/4$ retroreflector backfire configuration. 
The two transmitting antennas were mounted on a vertical configuration to minimize the advantages of a standard horizontal MIMO scheme.
Because of the different geometric spatial shapes of the phase fronts of the two beams, 
the field vectors of the $m=1$ twisted beam will result in anti-phase at two points on diametrically opposite sides of the OAM phase singularity at the centre of the two aligned beams and orthogonal to the propagation axis, whereas they will result in phase for an $m=0$ beam \cite{Thide&al:PRL:2007,Tamburini&al:APL:2011}. Therefore two simple standard linear-momentum interferometers, in anti-phase with respect to each others were used to discern between the twisted and untwisted beams \cite{Tamburini&al:NJP:2012}.
The antennas of the two interferometers were mounted at the same height from the floor and collinear on the same horizontal line. 

\begin{figure}
\centering - 
\includegraphics[width=8 cm]{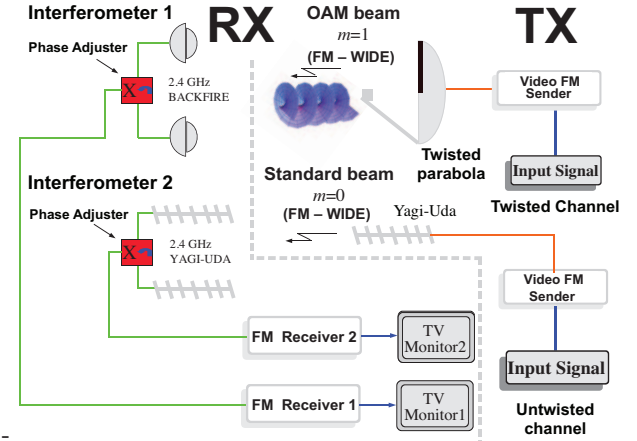}%
\caption{\label{fig:setup1}%
Schematic description of the setup of experiment 1 in the Foyer of ZKM. 
Two HD audiovideo channels were transmitted simultaneously in the same $2.414$ GHz frequency band, $27$ MHz bandwith, with the same horizontal polarization state. 
The twisted channel ($m=1$) was produced by a helicoidally deformed parabolic antenna and the untwisted  with dipole a standard antenna. 
At the receiving end, at a distance of $\sim 145 \lambda$ from the transmitting antennas the instantaneous EM fields of the $m=0$ and $m=1$ modes resulted in phase on one side of the central axis (azimuthal angle $\varphi=0$) and in anti-phase at the opposite side ($\varphi=\pi$). 
Therefore two independent standard linear-momentum interferometers one in phase with the $m=0$ beam and the other with the $m=1$ beam were used to receive simultaneously and uninterruptedly two transmissions. The boxes ``X'' indicates the relative phase adjuster present in each of the two interferometers.
}
\end{figure}

To minimize the possible cross-talking between the antennas of the two interferometers in each arm and antenna resonances, we used two couples of different types of antennas with the same gain.
The first interferometer was realized with two identical $26$~cm ($2.1 \lambda$) diameter $16$~dBi backfire antennas, connected together through a signal splitter/combiner and separated horizontally by a quantity $\sim 25 \lambda$.
The second interferometer was instead realized with a couple of $16$~dBi commercial off the shelf Yagi-Uda antennas, also connected together trough another signal splitter/combiner with a separation of $\sim 18 \lambda$ mounted on the same horizontal bar.
As shown in the scheme of Fig. \ref{fig:setup1}, in both receivers a phase tuner, in the form of a silver slit waveguide coupled to a Selenia signal circulator, was inserted into the path of the interferometers' arms.
In the first receiver we set the cursor of the slit waveguide so to retard the signal received by the first antenna  of half the wavelength antennas with respect to the second one and tune the receiver on the twisted channel.
In the second receiving line, instead, the slit waveguide was set so to have both the receiving antennas of the interferometer B in phase, canceling with disruptive interference the $m=1$ channel and revealing the untwisted $m=0$ one.
Once regulated the positions of the cursor in the waveguides and optimized the phase tuning, they remained fixed during the whole run. No additional electronic devices were used to adjust/correct the phase setting in the interferometers to compensate the phase errors occurring during the runs.
In this way we received simultaneously two channels in the same frequency band.
The averaged measured channel separation during Globale-Digitale with the digital channel was $32.3$ dB and standard deviation (std) of $0.8$. See Tab. \ref{tab:table1} for more details.

\textbf{Experiment 2:} the second experiment was part of a Science/Art installation located in the ``\textit{Infosphere}'' the of Globale--Digitale exhibition. The maximum output power was $0.1$ Watts per channel, same frequency and bandwidth as in experiment 1. 
In this link the reception of the two channels was performed by a single receiver at the end of an interferometer made with two $16$ dBi Yagi-Uda antennas. The selection between the two channels - twisted and untwisted - was obtained by dephasing the interferometer at the receiving end through the sliding one of the two receiving antennas, as in Ref.  \cite{Tamburini&al:NJP:2012}. 
The averaged channel separation measured with the digital channel was $22.4$ dB with $1.1$ std during the exhibition in $2015$. As for experiment 1, experiment 2 then moved in another room with similar conditions for the periodic tests in the following years. The results are reported in Tab \ref{tab:table1}.

\begin{figure}
\centering - 
\includegraphics[width=8 cm]{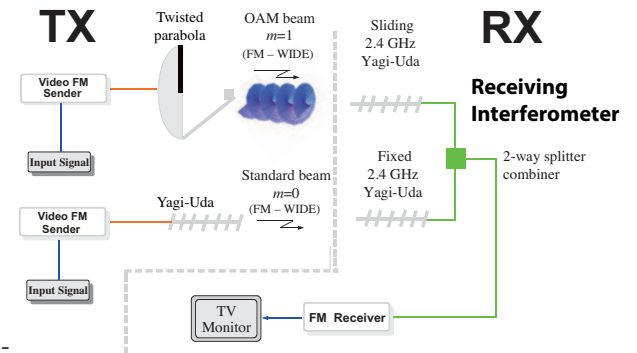}
\caption{\label{fig:setup2}%
Schematic description of the experimental setup 2 in the Infosphere of Globale at ZKM. 
The two high definition video channels were transmitted simultaneously in the same $2.414$ GHz frequency band with the same horizontal polarization state. 
At the receiving end, at a distance of $\sim 64 \lambda$ from the transmitting antennas the instantaneous EM fields of the $m=0$ and $m=1$ modes resulted in phase on one side of the central axis (azimuthal angle $\varphi=0$) and in anti-phase at the opposite side ($\varphi=\pi$). 
Therefore, the standard linear-momentum interferometer received the $m=1$ beam when there was a delay of $\lambda /2$ in the arms of the interferometer and the $m=0$ channel when the phase delay was null.
}
\end{figure}

\begin{table}[h!]
  \begin{center}
    \caption{Channel separation of experiment 1 and experiment 2 (dB) with their standard deviations and duration (months) $24$ hours per day.}
    \label{tab:table1}
    \begin{tabular}{lccccc} 
      \textbf{Year} & \textbf{exp. 1} & \textbf{std} & \textbf{exp. 2} & \textbf{std} & \textbf{duration}\\
      $ $ & dB & std & dB & std & months \\
      \hline
      2015 & 32.3 & 0.8 & 22.4 & 1.1 & 5\\
      2016 & 32.4 & 1.2 & 21.9 & 1.2 & 2\\
      2017 & 32.1 & 0.7 & 22.0 & 1.0 & 2\\
      2018 & 31.9 & 0.8 & 21.9 & 1.7 & 2\\
      2019 & 31.7 & 0.7 & 21.8 & 1.1 & 2\\
    \end{tabular}
  \end{center}
\end{table}

\begin{figure}
\centering - 
\includegraphics[width=8 cm]{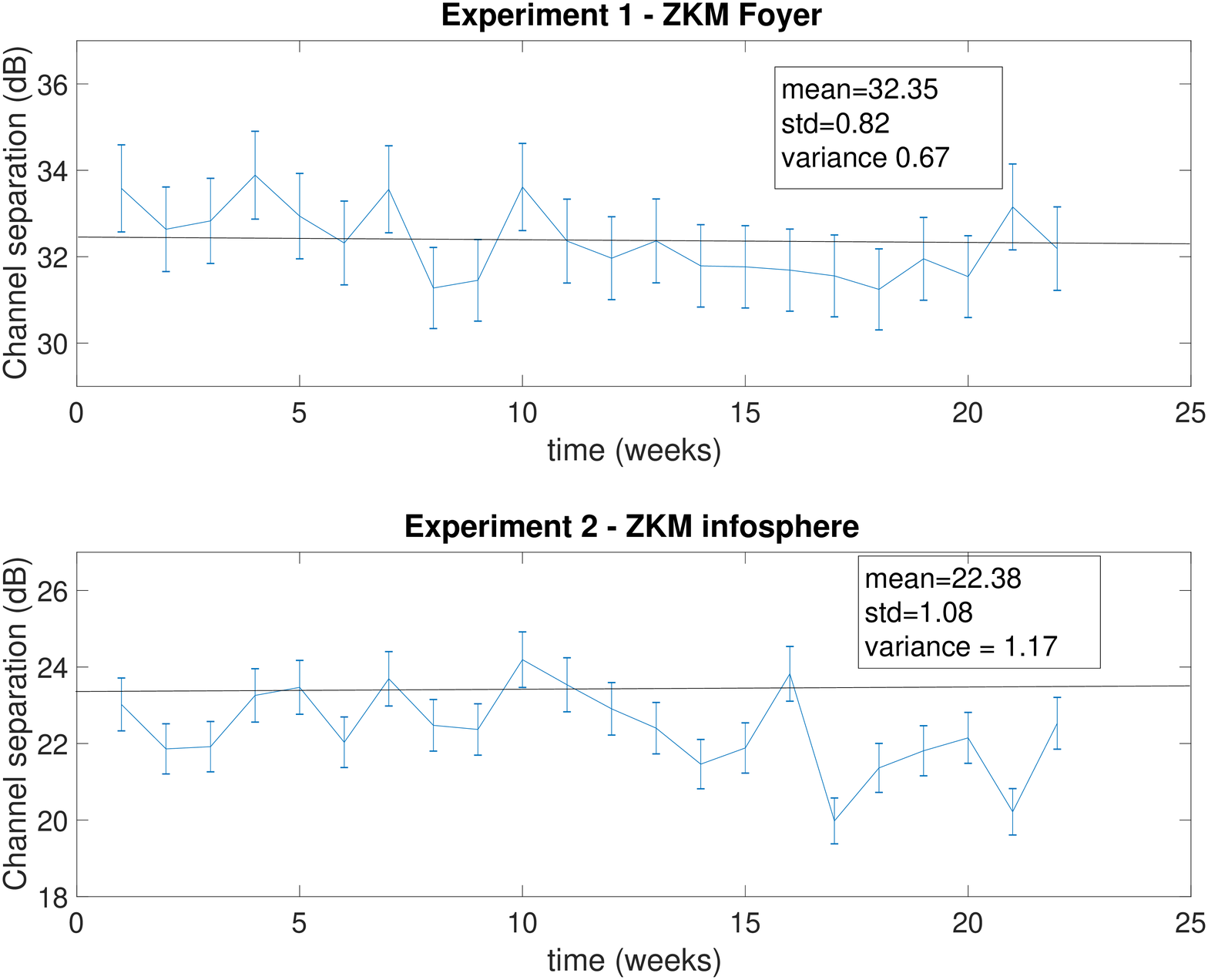}%
\caption{\label{runs}%
Result of the tests of channel separation with the digital channel of both links, experiment 1 and experiment 2 of the data acquisition in $2105$ during the Globale-Digitale exhibition. We notice that in the correspondence of conferences and events in the Foyer and in the Infosphere the channel separation were reaching the lower limit of the $20$ dB needed to transmit the two channels in the same frequency band.
}
\end{figure}

The experimental results reported here show that a multiple-channel transmission using OAM states is compatible and robust with respect to reflections, interference from a beam with different OAM value and does not need error correction code protocols like those used in digital channels.
More interestingly, the advantages offered by the use of digital multiplexing techniques, even those based on phase-coding such as phase shift keying (PSK), are preserved and can be used to further improve the channel capacity also in these complicated environmental situations as shown in Ref. \cite{Tamburini&al:preprint:2013a,triple,someda}. The channel separation values measured for the experiments 1 and 2 are drawn in Fig. \ref{runs} for the acquisition in $2015$ and in Tab. \ref{tab:table1} are reported the data of the acquisitions taken in all the five years of tests.

\subsection{Testing OAM multiplexing schemes with QPSK digital protocol}
Here we report the results of additional tests outdoor, in an urban setting, where both the twisted and untwisted channels experience controlled reflections and interferences with other disturbing signals that may overlap their frequency band.
The OAM beam was emitted by the twisted parabolic antenna, the untwisted mode was emitted by a $16$ dBi Yagi-Uda antenna.
At the receiver's side we did not use two independent interferometers as in experiment 1 because of logistic reasons. A single interferometer realized with the same $16$ dBi backfire antennas of experiment 1 was set up at the receiver's side and the tuning between the twisted and untwisted channels was realized with one of the phase tuners used in experiment 1. We superimposed on the same $2.414$~GHz frequency carrier the two collinear, horizontally-polarized radio beams, one twisted ($m=1$) and one untwisted ($m=0$).
As OAM radio links are compatible and robust with respect to digital multiplexing techniques, also with those based on phase coding such as the phase shift keying (PSK) coding protocol and QPSK, with very high coding rates \cite{Tamburini&al:preprint:2013a,triple,someda,Tamburini&al:preprint:2013a}, we used a quadrature phase shift coding (QPSK) DVB-S digital signal to measure with high precision the channel transmission capacity and channel insulation capacity of the twisted channel with respect to the untwisted channel.
QPSK is a digital phase encoding technique used in many telecommunications applications today, belonging to the class of PSK protocols. It employs, at any given time, four different phase states $\{ 01, 11, 10, 00 \}$ for the carrier. These four phase states correspond to $\{ 0, 90, 180, 270 \}$ degrees of relative phase shifts, respectively. For each time period, the phase can change once with constant amplitude. In this way, two bits of information are transmitted within each time slot.
This is the procedure used in our periodical tests for experiment 1 and experiment 2: the untwisted beam, transmitting at the same frequency and power as the twisted one, acts at all effects as a disturbing jamming signal.

\begin{figure}
\includegraphics[width=8 cm]{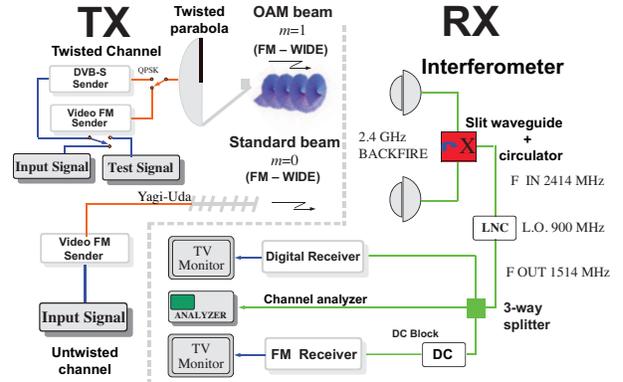}
\\
\caption{\label{fig:setupqpsk}%
\textbf{(Left: ) }Schematic description of the experimental setup for the channel test with digital protocols. 
The $m=1$ OAM beam carried a digital QPSK signal operating at a carrier frequency of $2.414$~GHz. 
The untwisted ($m=0$) signal at the same carrier frequency and with the same polarization was emitted by a standard Yagi-Uda antenna, fed by an FM $27$MHz wide band signal. 
An interferometer at the receiver's end equipped with a phase adjuster device was used to discriminate and test the two channels and the signal is downconverted to $1514$ MHz after being mixed with a local oscillator at $900$ MHz. Then the signal is divided and distributed to $3$ different paths: a digital receiver for QPSK, a channel analyzer and an analog FM receiver. 
In the worst conditions, because of the reflections, the modulation error ratio (MER) of the OAM $m=1$ QPSK channel superimposed to the $m=0$ one was $12$ dB and $20$ dB when the disturbing signal was switched off. The best value obtained without reflections was $50$ dB.}
\end{figure}

As shown and described in Fig.~\ref{fig:setupqpsk}, we probed the phases of the fields of the received signals at such points in a standard phase interferometric manner, using the two identical backfire antennas of experiment 1. 
The transmitted $m=1$ signal was encoded with a QPSK DVB-S protocol.
The digital transmitter used for the twisted ($m=1$) mode, a Microwave Link QPSK DVB-S transmitter for the $2.4$~GHz band was tuned to the usual frequency of $2.414$~GHz (free-space wavelength $\lambda=12.42$~cm) and transmitted  encoded video images at a rate of $11.5$~Megasymbols/s. 
We used a forward error correction of the FEC=$3/4$ type, i.e., after three bits transmitted, a fourth bit is added.
The $m=0$ untwisted channel used instead an analogue frequency modulation (FM) transmission, centered at $2.414$~GHz and $27$ MHz bandwidth, with $200$~mW maximum power and horizontal polarization.
The transmitting antenna was one of the $16$~dBi Yagi-Uda antennas used in the previous experiments.
We equalized the transmission channels by varying the output power of the two channels by inserting signal attenuators in the transmission line in order to obtain the same power at the receiving end for both the twisted and untwisted channels.
At the receiver's side, distant $80$~m ($644 \lambda$) from the transmitters, the signals were collected by an interferometer realized with the two $16$~dBi backfire antennas connected together through a signal combiner/splitter with a baseline of $2.6$ metres ($\sim 21 \lambda$). 
Both the antennas were placed about $1.5$~m ($\sim 12 \lambda$) above the reflecting ground.
As shown in Fig. \ref{fig:setupqpsk}, to select one of the two channels (twisted and untwisted), a phase tuner was placed in one of the interferometer arms.
The phase tuner was realized with a silver slit waveguide coupled to a Selenia signal circulator.
The received signal was routed to a signal analyzer to draw the frequency spectrum and intensity of the channels with a spectrum analyzer Hameg HMS3010 or to a 3-way power splitter where, after a down-conversion in frequency with a local oscillator (LO) of $900$ MHz to $1.514$ GHz was then split up into three different receiver chains where audiovideo channels were received, analyzed, and displayed in the monitors: the digital DVB-S digital chain with the QPSK constellation tester $(m=1)$, the scientific chain for measuring and testing in real time the analog ($m=0$) and digital ($m=1$) channels. Here the test-signal was down-converted to $48.25$ MHz and fed into one of the branches of the scientific channel to measure the received power level.
The analogue frequency modulation (FM) chain of the untwisted channel. In the latter chain we inserted a DC block that prevented the flow of audio and direct current (DC) frequencies and minimized the interference with the RF signals $(m=0)$.
The analyzer used was based on a Nokia-Hirschmann Receiver Analog Type HIT CSR $501$ satellite receiver. 
In this chain, the spectrum analyzer Hameg HMS3010 was inserted to measure the separation of the two channels, the signal levels, and the channel bandwidths. 

The modulation error ratio (MER) of the QPSK channel alone was larger than $20$~dB, with a bit error rate (BER) of $10^{-8}$ and a carrier-to-noise ratio C/N $>15$~dB, including the effect of the reflections mainly from the ground. The average background noise power in a $200$~MHz bandwidth was $-93$~dBm, peaking at $-85.6$~dBm at the centre frequency. The power of the received FM signal was measured by inserting the spectrum analyzer in the reception line and was found to be $-68.95$~dBm in a $27$~MHz wide transmission band.
When the FM analogue transmitter signal with the same carrier frequency was switched on, the ensuing interference on the digital twisted signal was quite small and caused the MER to vary from $9$ to $11$~dB, the BER from $10^{-3}$ to $10^{-5}$, and the C/N from $9$ to $12$~dB indicating that OAM states improve with their stable intrinsic phase properties future multiport schemes also in the presence of strong reflections.
In the absence of reflections we measured a channel separation of $50$ dB.

\section{Conclusions}
Radio transmissions with twisted waves carrying OAM are a robust tool for multiplexing schemes. Twisted waves are natural eigenmodes of the electromagnetic field that preserve in time their phase spatial profile when propagating in free space, as shown with the detection of twisted waves from the M87* black hole, located $56$ million light years far away from Earth \cite{tamburinimnras2019}.
We set up two indoor experiments working for long periods in time, the first run in $2015$ lasted for $5$ months, $24$ hours per day, and $2$ months per year, $24$ hours per day, in the following years until $2019$ to simulate real working conditions for future OAM links indoor where multiple reflections are present.
We periodically measured the separation between the two channels with digital channel techniques: experiment 1 had an averaged channel separation value of $32.1$ dB and experiment 2 $22.0$ dB.
Additional outdoor tests with controlled reflections were performed with a $4$-QAM digital transmission in the twisted channel. We obtained a maximum channel separation value of $50$ dB when the conditions of phase orthogonality were fulfilled and no reflections were present.
The results obtained clearly show that OAM can be used to improve the transmission capacity of common-use devices, and re-use efficiently the same frequency band, ensuring a good quality and stable channel separation for a long time.



\begin{thebibliography}{999}
\bibitem{mimo}
A. Sibille, C. Oestges, A. Zanella, \textit{MIMO: from Theory to Implementation.} Oxford, UK: Elsevier Academic Press, 2010.
\bibitem{fuschichnikitin}
W.I. Fushchich, A.G Nikitin, ``The complete sets of conservation laws for the electromagnetic field, '' 
\textit{Journal of Physics A: Mathematical and General}, vol. 25, pp. L231--L233, 1992.
\bibitem{Heitler1954}
W. Heitler, \textit{The Quantum Theory of Radiation.} Oxford, UK: Clarendon Press, The International Series of Monographs on Physics, 3 edn. Appendix 1, 1954.
\bibitem{Podolsky&Kunz1969}
B. Podolsky, K.S. Kunz, \textit{Fundamentals of Electrodynamics.} New York, NY, USA: Dekker,  chap.~V, 1969.
\bibitem{Eyges1972}
L. Eyges, \textit{The Classical Electromagnetic Field.} New York, NY, USA: Dover Publications, chap. 11, 1972.
\bibitem{Berestetskii&al1980}
V.B. Berestetskii, E.M. Lifschitz, L.P. Pitaevskii, \textit{Quantum Electrodynamics.} Oxford, UK: Pergamon Press,
Course of Theoretical Physics, vol.~ 4, chap.~ 1, 2nd edition, 1980.
\bibitem{Ribaric&Sustersic1990}
M. Ribari\v{c}, L. \v{S}u\v{s}ter\v{s}i\v{c}, \textit{Conservation Laws and Open Questions of Classical Electrodynamics.} London, UK: World Scientific, Singapore, 1990.
\bibitem{Mandel&Wolf1995}
L. Mandel, E. Wolf, \textit{Optical Coherence and Quantum Optics.} New York, NY, USA: Cambridge University Press, chap. 10, 1995.
\bibitem{Schwinger&al1998}
J. Schwinger, L.L. DeRaad Jr., K.A. Milton, W. Tsai, \textit{Classical Electrodynamics.} Reading, MA, USA: Perseus Books, chap. 3, 1998.
\bibitem{Jackson1998}
J.D. Jackson, \textit{Classical Electrodynamics} New York, NY, USA: Wiley, chap. 7 and 12, 3rd edition, 1998.
\bibitem{Allen&al2003}
L. Allen, S.M. Barnett, M.J. Padgett, \textit{Optical Angular Momentum.} Bristol, UK: IOP, 2003.
\bibitem{Rohrlich2007}
F. Rohrlich, \textit{Classical Charged Particles.} Singapore: World Scientific, chap. 7, 3rd edition, 2007.
\bibitem{Andrews2008}
D.L. Andrews, \textit{Structured Light and Its Applications: An Introduction to Phase-Structured Beams and Nanoscale Optical Forces.} Amsterdam, NL: Academic Press, 2008.
\bibitem{Torres&Torner2011}
J.P. Torres, L. Torner, \textit{Twisted Photons: Applications of Light With Orbital Angular Momentum.} Weinheim, DE: Wiley-Vch Verlag, John Wiley and Sons, 2011.
\bibitem{Yao&Padgett2011}
A.M. Yao, M.J. Padgett, ``Orbital angular momentum: origins, behavior and applications''. \textit{Adv. Opt. Photon.} vol. 3, pp. 161--204, 2011.
\bibitem{Thide2011}
B. Thid\'e, \textit{Electromagnetic Field Theory.} Mineola, NY, USA: Dover Publications, Inc. (In press), chap. 4, 2nd edition,  \url{http://www.plasma.uu.se/ CED/Book}.
\bibitem{Thide&al:PRL:2007}
B. Thid{\'e} 
\textit{et~al.} ``Utilization of photon orbital angular momentum in the low-frequency radio domain'', Phys. Rev. Lett. vol. 99,  pp. 087701(4), 2007.
\bibitem{Franke-Arnold&al:LPR:2008}
S. Franke-Arnold, L. Allen, M.J. Padgett, ``Advances in optical angular momentum,'' \textit{Laser, Photon. Rev.} vol.  2 ,  pp. 299--313, 2008.
\bibitem{Tamburini2011} 
F. Tamburini, B. Thid\'e, G. Molina-Terriza, G. Anzolin, ``Twisting of light around rotating black holes.'' \textit{Nature Phys.} vol. 7, pp. 195--197, 2011.
\bibitem{Tamburini&al:APL:2011}
F. Tamburini, E. Mari, B. Thid\'e, C. Barbieri, F. Romanato, ``Experimental verification of photon angular momentum and vorticity with radio techniques.'' \textit{Appl. Phys. Lett.} vol.  99, pp. 204102, 2011.
\bibitem{Tamburini&al:NJP:2012}
F. Tamburini, E. Mari, A. Sponselli, B. Thid\'e, A. Bianchini, F. Romanato, ``Encoding many channels on the same frequency through radio vorticity: first experimental test.'' \textit{New J.~Phys.} vol. 14, pp. 03301, 2012.
\bibitem{amradio2}
B. Thid\'e, F. Tamburini, H. Then, C.G. Someda, R.A. Ravanelli, ``The physics of angular momentum radio.'' arXiv:1410.4268 [physics.optics], 2014.
\bibitem{Tamburini&al:NJP:2012a}
F. Tamburini, E. Mari, A. Sponselli, B. Thid\'e, A. Bianchini, F. Romanato, ``Reply to comment: Encoding many channels on the same frequency through radio vorticity: first experimental test.'' \textit{New J.~Phys.} vol. 14, pp. 03301, 2012.
\bibitem{Andrews:Book:2008}
D.L. Andrews, \textit{Structured Light and Its Applications: An Introduction to Phase-Structured Beams and Nanoscale Optical Forces.} Amsterdam, NL: Academic Press, 2008.
\bibitem{oldoni}
M. Oldoni \textit{et al.} ``Space-division demultiplexing in orbital-angular-momentum-based MIMO radio systems.'' \textit{IEEE Transactions on Antennas and Propagation}, vol. 63, pp. 4582--4587, 2016.
\bibitem{Molina-Terriza&al:PRL:2002}
G. Molina-Terriza, J.P. Torres, L. Torner, ``Management of the angular momentum of light: Preparation of photons in multidimensional vector states of angular momentum'', \textit{Phys. Rev. Lett.} vol. 88, pp. 013601(4), 2002.
\bibitem{Mair&al:N:2001}
A. Mair, A. Vaziri, G. Weihs, A.  Zeilinger, ``Entanglement of the orbital angular momentum states of photons.'' \textit{Nature} vol. 412, pp. 313--316, 2001.  
\bibitem{Tamburini&al:preprint:2013a}
F. Tamburini {\it et al.} ``Experimental demonstration of free-space information transfer using phase modulated orbital angular momentum radio.'' arXiv:1302.2990 [physics.class-ph], 2013.
\bibitem{Gibson&al:OE:2004}
G. Gibson \textit{et~al.} ``Free-space information transfer using light beams carrying orbital angular momentum.'' \textit{Opt. Exp.} vol. 12, pp. 5448--5456, 2004.
\bibitem{Celechovsky&Bouchal:NJP:2007}
R. \v{C}elechovsk\`{y}, Z. Bouchal,  ``Optical implementation of the vortex information channel.'' \textit{New J.~Phys.}, vol. 9, pp. 328, 2007.
\bibitem{Wang&al:NPHO:2012}
J. Wang \textit{et~al.} ``Terabit free-space data transmission employing orbital angular momentum multiplexing.'' \textit{Nature Photon.}  vol. 6,  pp. 488--496, 2012.
\bibitem{bilotti18}
M. Barbuto, M.A. Miri, A. Al\'u, F. Bilotti, F. Toscano, ``Exploiting the Topological Robustness of Composite Vortices in Radiation Systems'', \textit{PIER}, vol. 162, pp. 39--50, 2018.
\bibitem{wileyIEEE}
B. Thid\'e, F. Tamburini, ``OAM Radio -- physical foundations and applications of electromagnetic orbital angular momentum in radio science and technology.'' in \textit{Electromagnetic Vortices: Wave Phenomena and Engineering Applications}, Wiley-IEEE Press, 2021.
\bibitem{amradio1}
B. Thid\'e, F. Tamburini, ``The physics of angular momentum radio'.' 2015 1st URSI Atlantic Radio Science Conference (URSI AT-RASC), Gran Canaria, Spain, 2015, \url{doi: 10.1109/URSI-AT-RASC.2015.7302908}.
\bibitem{tamburinimnras2019}
F. Tamburini, B. Thid\'e, M. Della Valle, ``Measurement of the spin of the M87 black hole from its observed twisted light.'' \textit{MNRAS Lett.}, vol. 492, pp. L22--L27, 2020. \url{https://doi.org/10.1093/mnrasl/slz176}.
\bibitem{FFT}
F. Tamburini, F. Feleppa, B. Thid\'e, ``Constraining the Generalized Uncertainty Principle with the light twisted by rotating black holes and M$87^{*}$.'' arXiv:2103.13750, 2021.
\bibitem{globale}
P. Weibel  \textit{et~al.} ``Globale Digitale (300 Years Karlsruhe -- 300 Days GLOBALE).'' 2015.
\url{https://zkm.de/en/blog/2015/05/300-years-karlsruhe-300-days-globale}.
\bibitem{Infosphere}
P. Weibel (Curator), D. Mille (Co-Curator), G. Bini (Co-Curator), ``Infosphere.''
\url{https://zkm.de/en/exhibition/2015/09/globale-Infosphere}.
\bibitem{einsteinsdream}
F. Tamburini, V. Boaga, B. Thid\'e, ``Beyond Einstein's Dream. Riding the Photons''  in \textit{Globale--Digitale}, artistic installation and Science and Art initiative (F. P. Grunert curator), 2015. \url{https://zkm.de/en/event/2015/09/beyond-einsteins-dream-riding-the-photons}.
\bibitem{triple}
F. Tamburini \textit{et al.} ``Tripling the capacity of a point-to-point radio link by using electromagnetic vortices.'' \textit{Radio Science}, vol. 50, pp. 501--508, 2015.
\bibitem{someda}
F. Spinello \textit{et al.} ``Radio channel multiplexing with superpositions of opposite-sign OAM modes.'' \textit{AEU-International Journal of Electronics and Communications}, vol. 70, pp. 990--997, 2016.
\end{thebibliography}
\end{document}